\def\den{\hbox{den}}

\def\tr{\hbox{tr}}
\def\ln{\ell{n}}
\documentstyle[a4,12pt]{article}
\begin{document}
\begin{titlepage} \vspace{0.2in} \begin{flushright}
MITH-96/1 \\ \end{flushright} \vspace*{1.5cm}
\begin{center} {\LARGE \bf  A Lattice Chiral Theory 
with Multifermion Couplings
\\} \vspace*{0.8cm}
{\bf She-Sheng Xue$^{(a)}$}\\ \vspace*{1cm}
INFN - Section of Milan, Via Celoria 16, Milan, Italy\\ \vspace*{1.8cm}
{\bf   Abstract  \\ } \end{center} \indent
 
Analyzing an $SU_L(2)\otimes U_R(1)$ chiral theory with multifermion couplings
on a lattice, we find a possible region in the phase space of multifermion
couplings, where no spontaneous symmetry breaking occurs, doublers are
decoupled as massive Dirac fermions consistently with the $SU_L(2)\otimes
U_R(1)$ chiral symmetry, the ``spectator'' fermion $\psi_R(x)$ is free mode,
whereas the normal mode of $\psi^i_L(x)$ is plausibly speculated to be chiral
in the continuum limit. This is not in agreement with the general belief of the
definite failure of theories so constructed. 

\vfill \begin{flushleft}  March, 1996   \vspace*{3cm} \\
\noindent{\rule[-.3cm]{5cm}{.02cm}} \\
\vspace*{0.2cm} \hspace*{0.5cm} ${}^{a)}$ 
E-mail address: xue@milano.infn.it\end{flushleft} \end{titlepage}
 
\noindent
{\bf 1.}\hspace*{0.3cm} 
It is a long standing problem to regularize chiral gauge theories on a lattice
and it seems that none of the methods proposed has been consistently and
completely demonstrated both to ensure that an asymptotically chiral gauge
theory in the continuum limit really exists and to provide a framework for
doing nonperturbative calculation in these theories \cite{p}. It is generally
believed that the constructions \cite{ep,xue} of chiral gauge theories on the
lattice with external multifermion couplings fail to give chiral gauged
fermions in the continuum limit for the reason\cite{gpr} that the theories so
constructed undergo spontaneous symmetry breaking and their phase structure is
similar to that of the Smit-Swift model\cite{ss}, which has been very carefully
studied and shown to fail. Nevertheless, we believe that further considerations
of constructing CGT on a lattice with external multifermion couplings and
careful studies of the spectrum in each phase of such a constructed theory are
necessary. In fact, we find a possible scaling region of defining continuum
chiral fermion in such a formulation of chiral gauge theories on the lattice. 

Let us consider the following fermion action of the $SU_L(2)$ CGT on a lattice
with two external multifermion couplings.
\begin{eqnarray}
S&\!=\!&{1\over 2a}\sum_x\left(\bar\psi^i_L(x)\gamma_\mu D^\mu_{ij}\psi^j_L(x)+
\bar\psi_R(x)\gamma_\mu\partial^\mu\psi_R(x)\right)\label{action}
\\
&\!+\!&
\sum_x\!\left(g_1\bar\psi^i_L(x)\!\cdot\!\psi_R(x)\bar\psi_R(x)\!\cdot\!\psi_L^i(x)
\!+\!g_2\bar\psi^i_L(x)\!\cdot\!\partial^2\psi_R(x)
\partial^2\bar\psi_R(x)\!\cdot\!\psi_L^i(x)\right),\nonumber
\end{eqnarray}
where ``$a$'' is the lattice spacing; $\psi^i_L$ ($i=1,2$) is an $SU_L(2)$
gauged doublet, $\psi_R$ is an $SU_L(2)$ singlet and both are two-component
Weyl fermions. The $\psi_R$ is treated as a ``spectator'' fermion. The second
multifermion coupling $g_2$, where 
\begin{equation}
\partial^2\psi_R(x)=\sum_\mu
\left[ \psi_R(x+\mu)+\psi_R(x-\mu)-2\psi_R(x)\right],
\nonumber
\end{equation}
is a dimension-10  
operator relevant only for doublers $p=\tilde p+\pi_A$,\footnote{The physical 
momentum $\tilde p\simeq 0$ 
and $\pi_A$ runs over fifteen lattice momenta $\pi_A\not=0$.}
but irrelevant for normal modes $p=\tilde p$ of the $\psi^i_L$ and 
$\psi_R$. In addition to the exact local $SU_L(2)$ chiral
gauge symmetry and the global chiral symmetry $SU_L(2)\otimes U_R(1)$, 
the action (\ref{action}) possesses a
$\psi_R$-shift-symmetry\cite{gp},
\begin{equation}
\psi_R(x) \rightarrow \psi_R(x)+{\rm const.},
\label{shift}
\end{equation}
when $g_1=0$. The chiral gauge interaction is supposed to be perturbative,
and we turn the gauge coupling off in the following discussions ($g=0$).

We consider the generating function $W(\eta)$, 
\begin{eqnarray}
W(\eta)&=&-\ln Z(\eta),
\nonumber\\
Z(\eta)&=&\int [d\psi_L^i d\psi_R]\exp\left(-S+\int_x\left(\bar\psi^i_L\eta_L^i+
\bar\eta_L^i\psi^i_L+\bar\psi_R\eta_R+
\bar\eta_R\psi_R\right)\right).
\label{part}
\end{eqnarray}
Then, we define the generating functional of one-particle irreducible vertices
(the effective action
$\Gamma(\psi'^i_L,\psi_R')$) as
the Legendre transform of $W(\eta)$
\begin{equation}
\Gamma(\psi'^i_L,\psi'_R)=W(\eta)-\int_x\left(\bar\psi'^i_L\eta_L^i+
\bar\eta_L^i\psi'^i_L+\bar\psi'_R\eta_R+
\bar\eta_R\psi'_R\right),\nonumber
\end{equation}
and with the relations
\begin{eqnarray}
\psi'^i_L(x)&=&\langle\psi^i_L(x)\rangle=-{\delta W\over\delta\bar\eta^i_L(x)
},\hskip1cm
\bar\psi'^i_L(x)=\langle\bar\psi^i_L(x)\rangle={\delta W\over\delta
\eta^i_L(x)},\label{pl}\\
\psi'_R(x)&=&\langle\psi_R(x)\rangle=-{\delta W\over\delta\bar\eta_R(x)},
\hskip1cm
\bar\psi'_R(x)=\langle\bar\psi_R(x)\rangle={\delta W\over\delta\eta_R(x)},
\label{pr}
\end{eqnarray}
in which the fermionic derivatives are left-derivatives, and
\begin{eqnarray}
\eta^i_L(x)&=&-{\delta\Gamma\over\delta\bar\psi'^i_L(x)},\hskip1cm
\bar\eta^i_L(x)={\delta\Gamma\over\delta\psi'^i_L(x)},\nonumber\\
\eta_R(x)&=&-{\delta\Gamma\over\delta\bar\psi'_R(x)},\hskip1cm
\bar\eta_R(x)={\delta\Gamma\over\delta\psi'_R(x)}.\label{er}
\end{eqnarray}
In eqs.(\ref{pl},\ref{pr}), the
$\langle\cdot\cdot\cdot\rangle$ indicates an expectation value with respect to 
the partition functional $Z(\eta)$ (\ref{part}).

We first derive the local Ward identity associated with the
$\psi_R$-shift-symmetry. Making the parameter $\epsilon$ to be spacetime
dependent, and varying the generating function (\ref{part}) according to the
transformation rules (\ref{shift}) for arbitrary $\epsilon(x)\not= 0$, we arrive
at 
\begin{equation}
\langle {1\over 2a}\gamma_\mu\partial^\mu\psi_R(x)
+g_1\bar\psi^i_L(x)\cdot\psi_R(x)\psi_L^i(x)
+g_2\partial^2\left(\bar\psi^i_L(x)\cdot
\partial^2\psi_R(x)\psi_L^i(x)\right)+\eta_R(x)\rangle=0.
\label{dw}
\end{equation}
Substituting (\ref{er}) into eq.(\ref{dw}), we obtain the Ward identity
corresponding to the $\psi_R$-shift-symmetry of the action (\ref{action}): 
\begin{equation}
{1\over 2a}\gamma_\mu\partial^\mu\psi^\prime_R(x)
+g_1\langle\bar\psi^i_L(x)\!\cdot\psi_R(x)\psi_L^i(x)\rangle
\!+g_2\!\langle\partial^2\!\left(\bar\psi^i_L(x)\!\cdot\!
\partial^2\psi_R(x)\psi_L^i(x)\right)\rangle-{\delta\Gamma\over\delta\bar
\psi'_R(x)}=0.
\label{w}
\end{equation}
Based on this Ward identity (\ref{w}), one can get all one-particle irreducible
vertices $\Gamma^{(n)}_R$ containing at least one external $\psi_R$. 

Taking functional derivatives of eq.(\ref{w}) with respect to appropriate
``prime'' fields (\ref{pl},\ref{pr}) and then putting external sources $\eta=0$, one can derive:
\begin{eqnarray}
\int_xe^{-ipx}
{\delta^{(2)}\Gamma\over\delta\psi'_R(x)\delta\bar\psi'_R(0)}\!&=&\!{i\over a}
\gamma_\mu\sin (p^\mu a),\label{free}\\
\int_xe^{-ipx}
{\delta^{(2)}\Gamma\over\delta\psi'^i_L(x)\delta\bar\psi'_R(0)}\!&=&\!
{1\over2}\Sigma^i(p)\nonumber\\
\!&=&\!g_1\langle\bar\psi^i_L(0)\!\cdot\!\psi_R(0)\rangle_\circ
\!+\!2g_2w(p)\langle\bar\psi^i_L(0)\cdot
\partial^2\psi_R(0)\rangle_\circ,
\label{ws2}
\end{eqnarray}
where the
$\langle\cdot\cdot\cdot\rangle_\circ$ indicates an expectation value with 
respect to the partition functional $Z(\eta)$ (\ref{part}) without external 
sources $(\eta=0)$. In addition, one can derive the four-point vertex,
\begin{equation}
\int_{xyz}e^{-iyq-ixp-izp'}
{\delta^{(4)}\Gamma\over\delta\psi'^i_L(0)\delta\bar\psi'^i_L(y)\delta\psi'_R(z)
\delta\bar\psi'_R(x)}\!=\! g_1\!+\!4g_2w(p+{q\over 2})w(p'+{q\over 2}),
\label{4p}
\end{equation}
where $p+{q\over 2}$ and $p'+{q\over 2}$ are momenta of the $\psi_R$ field.
In eqs.(\ref{ws2},\ref{4p}), $w(p)$ is the well-known Wilson factor
\begin{equation}
w(p)=\sum_\mu\left(1-\cos(p_\mu)\right),
\nonumber
\end{equation}
and all momenta are scaled to be dimensionless.
All other one-particle
irreducible vertices $\Gamma_R^{(n)}=0 (n>4)$ identically. When 
$g_1=0$, we find (i) eq.(\ref{free}) shows that the $\psi_R(x)$ is free field;
(ii) eq.(\ref{4p}) for
the normal mode of the $\psi_R$ are vanishing at least $O((ma)^2)$, where $m$
is the scale of the continuum limit.  This may indicate
that when $g_1=0$, the normal mode of the $\psi_R$ completely decouples and 
dose not form any bound states with other modes.   

\vskip0.7cm
\noindent
{\bf 2.}\hspace*{0.3cm} 
Our goal is to seek a possible regime, where an undoubled $SU_L(2)$-chiral
gauged fermion content is exhibited in the continuum limit in the phase space
$(g_1,g_2,g)$, where ``$g$'' is the gauge coupling,  regarded to be a truly
small perturbation $g\rightarrow 0$ at the scale of the continuum limit we
consider. In the weak coupling limit, $g_1\ll 1$ and $g_2\ll 1$ (indicated 1 in
fig.1), the action (\ref{action}) defines an $SU_L(2)\otimes U_R(1)$ chiral
continuum theory with a doubled and weakly interacting fermion spectrum that is
not the continuum theory we seek. 

Let us consider the phase of spontaneous symmetry breaking in the weak-coupling
$g_1,g_2$ limit. Based on the analysis of large-$N_f$ ($N_f$ is an extra
fermion index, e.g.,~color, $N_c$) weak coupling expansion, we show that the
multifermion couplings in the action (\ref{action}) undergo Nambu-Jona Lasinio
(NJL) spontaneous chiral-symmetry breaking\cite{njl}. In this symmetry breaking
phase (indicated 2 in fig.1), the $SU_L(2)\otimes U_R(1)$-chiral symmetry is
violated by 
\begin{equation}
{1\over2}\Sigma^i(p)=g_1\int d^4x e^{-ipx}
\langle\bar\psi^i_L(0)\cdot\psi_R(x)\rangle_\circ\not=0.
\nonumber
\end{equation}
Assuming that the symmetry breaking takes place in the direction 1 in the
2-dimensional space of the $SU_L(2)$-chiral symmetry ($\Sigma^1(p)\not=0, 
\Sigma^2(p)=0$), 
one finds the following fermion spectrum that contains a doubled Weyl fermion
$\psi^2_L(x)$ and a undoubled Dirac fermion made by the Weyl fermions
$\psi^1_L(x)$ and $\psi_R(x)$. The propagators of these fermions can be written
as, 
\begin{eqnarray}
S_{b1}^{-1}(p)&=&{i\over a}\sum_\mu\gamma_\mu\sin p_\mu Z_2(p)P_L
+{i\over a}\sum_\mu\gamma_\mu \sin p^\mu P_R+\Sigma^1(p)\label{sb1}
\\
S_{b2}^{-1}(p)&=&{i\over a}\sum_\mu\gamma_\mu\sin p_\mu Z_2(p)P_L.
\label{sb2}
\end{eqnarray}
The $SU_L(2)\otimes U_R(1)$ chiral symmetry is realized to be $U_L(1)\otimes
U(1)$ with three Goldstone modes and a massive Higgs mode that are not
presented in this short report.

Owing to the four-fermion interaction vertex (\ref{4p}), 
the fermion self-energy function $\Sigma^1(p)$ in eqs.(\ref{ws2}) and 
(\ref{sb1}) is given by the NJL gap-equation in the large-$N_f$ weak coupling
expansion ($N_f\rightarrow\infty$)
\begin{equation}
\Sigma^1(p)=4\int_q{\Sigma^1(q)\over\den(q)}\left(\tilde g_1+4\tilde g_2w(p)w(q)
\right)
\label{se}
\end{equation}
where 
\begin{eqnarray}
\int_q&\equiv& \int_\pi^\pi{d^4q\over (2\pi)^4}\nonumber\\
\den(q)&\equiv&
\sum_\rho\sin^2q_\rho +(\Sigma^1(q)a)^2\nonumber\\
\tilde g_1&\equiv& g_1N_fa^2,\hskip1cm \tilde
g_2\equiv g_2N_fa^2.\nonumber
\end{eqnarray}
We adopt the paramatrization\cite{gpr}
\begin{equation}
\Sigma^1(p)=\Sigma^1(0)+\tilde g_2 v^1w(p), \hskip1cm 
\Sigma^1(0) =\rho v^1,\label{para}
\end{equation}
where $\rho$ depends only on
couplings $\tilde g_1, \tilde g_2$, and $v^1$ plays a role as the v.e.v.
violating $SU_L(2)\otimes U_R(1)$-chiral symmetry. We can solve the 
gap-equation (\ref{se}) by using this paramatrization (\ref{para}).
For $v^1=O({1\over a})$, one obtains 
\begin{equation} 
\rho={\tilde g_1\tilde
g_2 I_1\over 1-\tilde g_1 I_\circ}; \hskip1cm \rho={1-4\tilde g_2 I_2\over
4 I_1},\label{rho} 
\end{equation} 
where the functions $I_n(v^1), (n=0,1,2)$, are defined as
\begin{equation}
I_n(v^1)=4\int_q{w^n(q)\over\sum_\rho\sin^2q_\rho +(\Sigma^1(q)a)^2}.\nonumber
\end{equation}
Eq.(\ref{rho}) leads to a crucial result:
\begin{equation}
\tilde g_1=0,\hskip1cm \rho=0\hskip0.5cm and\hskip0.5cm 
\Sigma^1(0)=0,\label{o}
\end{equation}
this is due to eq.(\ref{ws2}) resulted from the Ward identity (\ref{w}). 
This means that on the line $g_1$=0,
normal modes $(p=\tilde p\simeq 0)$ of the $\psi^1_L$ and $\psi_R$ are
massless and their 15 doublers $p=\tilde p+\pi_A$ acquire chiral-variant masses
\begin{equation}
\Sigma^1(p)=\tilde g_2 v^1w(p)
\nonumber
\end{equation}
through the multifermion coupling $g_2$ {\it only}. In this case 
($g_1=0$), the gap-equation is then given by eq.(\ref{rho}) for $\rho=0$,
\begin{equation}
1-4\tilde g_2 I_2(v^1)=0,\hskip0.3cm i.e.\hskip0.3cm
1=16\tilde g_2\int_q{w^2(q)\over\sum_\rho\sin^2q_\rho +(\tilde g_2v^1w(q)a)^2}.
\nonumber
\end{equation}

As $v^1\rightarrow 0$, eq.(\ref{rho}) gives a
critical line $\tilde g^c_1 (\tilde g^c_2)$ 
\begin{equation}
\tilde g^c_1={1-4\tilde g_2^c I_2(0)\over4\tilde g_2^cI_1^2(0)
+I_\circ(0)
-4\tilde g_2^cI_\circ(0) I_2(0)},\nonumber
\end{equation} 
of characterizing NJL spontaneous
chiral symmetry breaking. With $I_\circ(0)=2.48, I_1(0)=4I_\circ(0)$ and $
I_2(0)=20I_\circ(0)-4$, the critical points are given by:
\begin{equation}
\tilde g^c_1=0.4,\hskip0.5cm \tilde g^c_2=0;\hskip1.5cm 
\tilde g^c_1=0,\hskip0.5cm \tilde g^c_2=0.0055,\label{wcri}
\end{equation}
as indicated 2 in fig.1. These critical values are
sufficiently small evenfor $N_f=1$. 

As for the wave function renormalization $Z_2(p)$ in eqs.(\ref{sb1},\ref{sb2}),
it depends on the dynamics of the left-handed Weyl fermion $\psi_L^i$ in this
region. In large-$N_f$ calculation at weak couplings, we are able to evaluate
this function $Z_2(p)$. The result is not presented in this short report. 

This broken phase cannot be a candidate for a real chiral
gauge theory (e.g.,~the Standard Model) for the reasons that $(i)$ $\psi^2_L$ is
doubled (\ref{sb2}); $(ii)$ the spontaneous symmetry breakdown of the
$SU_L(2)$-chiral symmetry is caused by the hard breaking Wilson
term\cite{wilson} (\ref{sb1})(dimension-5 operator), which must contribute
the intermediate gauge boson masses through the perturbative
gauge interaction and disposal of Goldstone modes. The intermediate gauge boson
masses turn out to be $O({1\over a})$. This, however, is
phenomenologically unacceptable. 

\vskip0.7cm
\noindent
{\bf 3.}\hspace*{0.3cm} 
We turn to the strong coupling region, where $g_1(g_2)$ are sufficiently larger
than certain critical values: 
\begin{equation}
g_1(g_2)\gg g_1^c(g_2^c)
\label{threshold}
\end{equation}
(indicated 3 in fig.1). Analogously to the analysis and discussions of Eichten
and Preskill (EP) \cite{ep}, we can show that the $\psi^i_L$ and $\psi_R$ in
(\ref{action}) are bound up to form the composite Weyl fermions (three-fermion
bound states)
\begin{equation}
\Psi^n_L={1\over2a}(\bar\psi^i_L\cdot\psi_R)\psi^i_L
\label{cl}
\end{equation}
(left-handed $SU_L(2)$-neutral) 
and 
\begin{equation}
\Psi_R^i={1\over2a}(\bar\psi_R\cdot\psi^i_L)\psi_R
\label{cr}
\end{equation}
(right-handed $SU_L(2)$-charged). These three-fermion bound states (\ref{cl},\ref{cr})
respectively pair up with the
$\bar\psi_R$ and $\bar\psi_L^i$ to be massive, neutral $\Psi_n$ and charged
$\Psi_c^i$ Dirac modes 
\begin{equation}
\Psi_n=(\Psi_L^n, \psi_R);\hskip1cm\Psi^i_c=(\psi_L^i, \Psi^i_R),
\label{di}
\end{equation}
consistently with the $SU_L(2)\otimes U_R(1)$ chiral
symmetry. The propagators of these Dirac fermions are given by\footnote{I thank 
Y.~Shamir for discussions on these propagators.}
\begin{equation}
\langle\Psi^i_c(0)\bar\Psi^j_c(x)\rangle\!=\!
\langle\psi^i_L(0)\bar\psi^j_L(x)\rangle\!+\!\langle\Psi^i_R(0)\bar\psi^i_L(x)
\rangle\!+\!\langle\psi^i_L(0)\bar\Psi^j_R(x)\rangle\!+\!\langle\Psi^i_R(0)
\bar\Psi^j_R(x)\rangle,
\label{dc}
\end{equation}
and
\begin{equation}
\langle\Psi_n(0)\bar\Psi_n(x)\rangle \!=\!
\langle\Psi^n_L(0)\bar\Psi^n_L(x)\rangle\!+\!\langle\Psi^n_L(0)\bar\psi_R(x)
\rangle\!+\!\langle\psi_R(0)\bar\Psi^n_L(x)\rangle\!+\!\langle\psi_R(0)
\bar\psi_R(x)\rangle,
\label{dn}
\end{equation}
which we need to compute. These fermions are, in general, massive. But, {\it
a priori}, we cannot exclude the possibility of massless composite modes. This
is, as we will see, not what we desire.

In order to compute these fermion propagators, we use strong multifermion 
coupling, 
\begin{equation}
g_1\gg 1,\hskip1cm g_2=0,
\label{strong1}
\end{equation}
expansion in the powers of $({1\over g_1})$.
We obtain the following recursion relations \cite{we} in the lowest 
nontrivial order,
\begin{eqnarray}
S^{ij}_{LL}(x)&=&{1\over g_1}\left({1\over 2a
}\right)^3\sum^\dagger_\mu S^{ij}_{ML}(x+\mu)\gamma_\mu,\label{re1}\\
S^{ij}_{ML}(x)&=&{\delta(x)\delta_{ij}\over 2g_1}
+{1\over g_1}\left({1\over 2a
}\right)\sum^\dagger_\mu S^{ij}_{LL}(x+\mu)\gamma_\mu,
\label{re2}\\
S^{ij}_{MM}(x)&=&{1\over g_1}\left({1\over 2a
}\right)\sum^\dagger_\mu 
\gamma_\mu\gamma_\circ S^{ij\dagger}_{ML}(x+\mu)\gamma_\circ,
\label{re3}
\end{eqnarray}
where definition
\begin{equation}
\sum^\dagger_\mu f(x)\equiv \sum_\mu\left(f(x+\mu)-f(x-\mu)\right),
\nonumber
\end{equation}
and two-point functions are 
\begin{eqnarray}
S^{ij}_{LL}(x)&\equiv&\langle\psi^i_L(0),\bar\psi^j_L(x)\rangle,\label{sll}\\
S^{ij}_{ML}(x)&\equiv&\langle\psi^i_L(0),[\bar\psi^j_L(x)\cdot\psi_R(x)]
\bar\psi_R(x)\rangle,
\label{sml}\\
S^{ij}_{MM}(x)&\equiv&\langle[\bar\psi_R(0)\cdot\psi^i_L(0)]
\psi_R(0),[\bar\psi^j_L(x)\cdot\psi_R(x)]
\bar\psi_R(x)\rangle.
\label{smm}
\end{eqnarray}
in the propagator of charged Dirac fermion (\ref{dc}). As for the neutral
fermion propagator (\ref{dn}), results are analogous to (\ref{re1},\ref{re2},
\ref{re3}). Thus, we calculate the propagators of neutral and 
charged Dirac modes to be
\begin{eqnarray}
S_n(p)\!&=\!&{{i\over a}\sum_\mu\gamma_\mu\sin p^\mu +M_1\over
{1\over a^2}\sum_\rho\sin^2p_\rho+M_1^2}
\label{sn11}\\
S_c(p)_{ij}\!&=\!&\delta_{ij}{{i\over a}\sum_\mu\gamma_\mu\sin p^\mu +M_1
\over {1\over a^2}\sum_\rho\sin^2p_\rho+M_1^2},
\label{sc11}\\
M_1&=&2ag_1.\label{m1}
\end{eqnarray}
This spectrum, which consists of sixteen modes of neutral Dirac fermion and 
sixteen modes of
charged Dirac fermion, is massive (degenerate) and vector-like consistently 
with the $SU_L(2)\otimes U_R(1)$ chiral symmetry.

Similar strong coupling expansion in the powers of $({1\over g_2})$ can be 
performed in the case that
\begin{equation}
g_2\gg 1,\hskip1cm g_1=0.\nonumber
\end{equation}
We obtain\cite{we} the results that are analogous to 
eqs.(\ref{sn11}) and (\ref{sc11}),
\begin{eqnarray}
S_n(p)\!&=\!&{{i\over a}\sum_\mu\gamma_\mu\sin p^\mu +M_2(p)\over
{1\over a^2}\sum_\rho\sin^2p_\rho+M_2^2(p)}
\label{sn1}\\
S_c(p)_{ij}\!&=\!&\delta_{ij}{{i\over a}\sum_\mu\gamma_\mu\sin p^\mu +
M_2(p)
\over {1\over a^2}\sum_\rho\sin^2p_\rho+M_2^2(p)},
\label{sc1}\\
M_2(p)&=&8ag_2w(p)\hskip1cm p\not=\tilde p.\label{m2}
\end{eqnarray}
Instead of eq.(\ref{m1}), the chiral-invariant masses $M_2(p)$ (\ref{m2}) of
doublers are not degenerate. It is very important to note that these equations
(\ref{sn1},\ref{sc1}) are not valid for the normal modes $p=\tilde p$. This
spectrum, which consists of fifteen doublers of neutral Dirac fermion and 
fifteen doublers of
charged Dirac fermion, is massive and vector-like consistently with the
$SU_L(2)\otimes U_R(1)$ chiral symmetry. As for the normal modes $(p=\tilde
p)$ of these composite Dirac fermions, their propagators in the strong coupling region are to our knowledge still
lacking. On the basis of the following discussions in next section, we might
expect that these normal modes of three-fermion bound states (\ref{cl}) and
(\ref{cr}) have not been bound yet and thus the spectrum of normal modes is
chiral, provided the multifermion couplings (\ref{4p}) are momentum-dependent
and not strong enough in a certain region of the phase diagram.

\vskip0.7cm
\noindent
{\bf 4.}\hspace*{0.3cm} 
The critical value $g^c_1(g^c_2)$ (\ref{threshold})
that we have mentioned in the beginning of
section 3 can be determined by considering \cite{gpr} the propagator
$G^{ij}(q)$ of a complex composite field ${\cal A}^i$, 
\begin{equation}
G^{ij}(\tilde q)=\int d^4xe^{-i\tilde qx}\langle {\cal A}^i(0)
{\cal A}^{\dagger j}(x)
\rangle_\circ,
\hskip1cm{\cal A}^i=\bar\psi_R\cdot\psi^i_L.\nonumber
\end{equation}
The real and imaginary parts of ${\cal A}^i(x)$ are
four composite scalars ($i=1,2$),
\begin{eqnarray}
A_1^i&=&{1\over2}(\bar\psi^i_L\cdot\psi_R+\bar\psi_R\cdot
\psi^i_L)\nonumber\\
A_2^i&=&{i\over2}(\bar\psi^i_L\cdot\psi_R-\bar\psi_R\cdot\psi^i_L).
\nonumber
\end{eqnarray}
Again using the strong coupling (\ref{strong1}) expansion in the powers of
(${1\over g_1}$) and we obtain the
recursion relation in the lowest nontrivial order, 
\begin{equation}
G^{ij}(\tilde q)={\delta_{ij}\over g_1}+\left(
{1\over 2a^2}\right){1\over g_1}\sum_{\pm\mu}\cos 
\tilde q_\mu G^{ij}(\tilde q).
\nonumber
\end{equation}
As a result, we find these four massive composite scalar modes,
\begin{equation}
G^{ij}(\tilde q)= {4\delta_{ij}\over {4\over a^2}\sum_\mu\sin^2{\tilde 
q_\mu\over 2}
+\mu^2};\hskip0.5cm \mu^2= 4\left(g_1-{2\over a^2}\right),
\label{mas}
\end{equation}
which are degenerate owing to the exact $SU_L(2)\otimes U_R(1)$-chiral
symmetry. Thus, $\mu^2{\cal A}^i{\cal A}^{\dagger i}$ gives rise to
a quadratic mass term of the composite scalar field ${\cal A}^i$ in the 
effective Lagrangian.  We assume that the one particle irreducible 
vertex ${\cal A}^j
{\cal A}^{\dagger j}{\cal A}^i{\cal A}^{\dagger i}$ is positively definite 
and the 
energy of ground state is bound from bellow. A spontaneous symmetry breaking 
$SU(2)\rightarrow U(1)$ occurs, where
$\mu^2>0$ turns to $\mu^2<0$. Eq.(\ref{mas}) for $\mu^2=0$ gives rise to the
critical point:
\begin{equation}
g_1^ca^2=2,\hskip1cm g_2=0,
\label{scri}
\end{equation}
(as indicated in Fig.1) where a phase transition takes place
between the NJL symmetry breaking phase and the EP symmetric phase. 

The second multifermion coupling $4g_2w(p+{\tilde q\over2}) w(p'+{\tilde
q\over2})$ in (\ref{4p}) gives different contributions to the effective value
of $g_1$ at large distance for sixteen modes of the $\psi_L^i$ and $\psi_R$ in
the action (\ref{action}). We should not doubt that the critical lines 
$g_1^c(g_2^c)$
(the thresholds of forming three-fermion bound states) should depend on sixteen
modes of the $\psi_L^i$ and $\psi_R$. These critical lines can be qualitatively
determined in the following considerations.  
Substituting the coupling $g_1$ in eq.(\ref{mas}) by the effective coupling
(\ref{4p}), one gets
\begin{equation}
\mu^2= 4\left(g_1+4g_2w(p+{\tilde q\over2}) w(p'+{\tilde
q\over2})-{2\over a^2}\right).\label{threshold2}
\end{equation}
Let us consider the
multifermion couplings of each mode ``$p$'' of the $\psi_L^i$ and $\psi_R$,
namely, we set $p=p', q=\tilde q\ll 1$ in the four-point vertex (\ref{4p}).
one gets
\begin{equation}
\mu^2= 4\left(g_1+4g_2w^2(p)-{2\over a^2}\right).\label{threshold2'}
\end{equation}
Thus, $\mu^2=0$ gives rise to the critical
lines:
\begin{equation}
g_1^ca^2=2, g_2=0;\hskip0.5cm g_1=0, a^2g_2^{c,b}=0.008,
\label{threshold1}
\end{equation}
where the first binding threshold of the doubler $p=(\pi,\pi,\pi,\pi)$ is, 
and 
\begin{equation}
g_1^ca^2=2, g_2=0;\hskip0.5cm g_1=0, a^2g_2^{c,a}=0.124,
\label{thresholdl}
\end{equation}
where the last binding threshold of the doublers $p=(\pi,0,0,0)$ is. Inbetween
(indicated 4 in fig.1) there are the binding thresholds of the doublers
$p=(\pi,\pi,0,0)$ and $p=(\pi,\pi,\pi,0)$ in eq.(\ref{threshold2'}), and the binding thresholds of the
different doublers $p\not=p'$ in eq.(\ref{threshold2}). Above $g_1^{c,a}$ {\it
all} doublers are supposed to be bound, as indicated 5 in fig.1. As for the
normal modes of the $\psi^i_L$ and $\psi_R$, when $g_1\ll 1$, the multifermion
coupling (\ref{4p}), $\Gamma^{(4)}=g_1+4g_2w^2(\tilde p)$, is supposed to be no
longer strong enough to form the bound states
$(\bar\psi^i_L\cdot\psi_R)\psi^i_L$ and $(\bar\psi_R\cdot\psi^i_L)\psi_R$ unless
$a^2g_2 \rightarrow\infty$. It is conceivable that the critical line for normal
modes, which is given by eq.(\ref{threshold2'}) for $\tilde p=ma\ll 0$,
\begin{equation}
g_1+ag_2O((ma)^4)-{1\over 2a^2}=0,
\nonumber
\end{equation}
analytically continues to the limit 
\begin{equation}
g_2^{c,\infty}\rightarrow \infty,\hskip0.5cm g_1\rightarrow 0.
\nonumber
\end{equation}

\vskip0.7cm
\noindent
{\bf 5.}\hspace*{0.3cm}
We must confess that the description of momentum dependence of the threshold
should not certainly be considered a rigorous demonstration. Nevertheless, we
can see, as expected in ref.\cite{ep}, several wedges open up as $g_1, g_2$
increase in the NJL phase (indicated 5 in fig.1), inbetween the critical lines
along which bound states of normal modes and doublers of the $\psi^i_L$ and
$\psi_R$ respectively approach their thresholds. In the initial part of the NJL
phase, the normal modes and doublers of the $\psi^i_L$ and $\psi_R$ undergo the
NJL phenomenon and contribute to eqs.(\ref{sb1},\ref{sb2}) as discussed in
section 2. As $g_1,g_2$ increase, all these modes, one after another, gradually
disassociate from the NJL phenomenon and no longer contribute to
eqs.(\ref{sb1},\ref{sb2}). Instead, they turn to associate with the EP
phenomenon and contribute to eqs.(\ref{sn1},\ref{sc1}) and
eqs.(\ref{sn11},\ref{sc11}).  The first and last doublers of the $\psi^i_L$ and
$\psi_R$ making this transition are $p=(\pi,\pi,\pi,\pi)$ and $p=(\pi,0,0,0)$
respectively. At the end of this sequence, normal modes ($p=\tilde p)$ make
this transition, due to the fact that they possess the different effective
multifermion coupling $\Gamma^{(4)}=g_1+4g_2w^2(p)$. 

Had these critical lines separated the two symmetric phases, (strong couplings
and the weak coupling symmetric phases) we would have found a threshold over
which all doublers of the $\psi_L^i$ and $\psi_R$ decouple by acquiring chiral
invariant masses (\ref{m1},\ref{m2}) and normal modes of the $\psi_L^i$ and
$\psi_R$ remain massless and free, and we might obtain a theory of massless free
chiral fermions \cite{ep}. However, this is not real case \cite{gpr}. As has
been seen in eq.(\ref{mas}), turning $\mu^2>0$ to $\mu^2<0$ indicates a phase
transition between the strong coupling symmetric phase to the spontaneous
chiral symmetry breaking phase, which separates the strong coupling and 
weak coupling symmetric phases. As indicated in Fig.~1, this can be clearly 
seen in eq.(\ref{wcri}) and eqs.(\ref{threshold1}).

The possible resolution of this undesired situation is that we can find a
region in which the doublers of the $\psi^i_L$ and $\psi_R$ have formed bound
states $(\bar\psi_R\cdot\psi^i_L)\psi_R$ and
$(\bar\psi^i_L\cdot\psi_R)\psi^i_L$ via the EP phenomenon, while the normal
modes of the $\psi^i_L$ and $\psi_R$ have neither formed such bound states yet
and nor are they associated with the NJL-phenomenon. 

Let us try to find whether there is a such resolution. Within the last wedge
(indicated 5 in fig.1) between two the thresholds $g_2^{c,a}$ and
$g_2^{c,\infty}$, all doublers of the $\psi^i_L$ and $\psi_R$ are bound to be
Dirac fermions that acquire chiral-invariant masses and decouple (considering
that eqs.(\ref{sn1},\ref{sc1}) are the propagators for doublers $p=\tilde
p+\pi_A$ {\it only}), and the normal modes of the $\psi^i_L$ and $\psi_R$ are
supposed not yet to be bound as Dirac fermions (we have not rigorously proved
this point). Thus, we have the undoubled low-energy spectrum that involves only
the normal modes of the $\psi^i_L$ and $\psi_R$. However, because of the
multifermion coupling $g_1\not= 0$, these normal modes of $\psi^i_L$ and
$\psi_R$ still remain in the NJL broken phase, the $SU_L(2)\otimes
U_R(1)$-chiral symmetry is violated by $\Sigma^1(0)=\rho v^1\not=0$, 
to which only
normal modes contribute. The propagators of the normal modes in this wedge
should be the same as eqs.(\ref{sb1},\ref{sb2}) for $p=\tilde p$
\begin{eqnarray}
S_{b1}^{-1}(\tilde p)&=&i\sum_\mu\gamma_\mu\tilde p_\mu Z_2(\tilde p)P_L
+i\sum_\mu\gamma_\mu \tilde p^\mu P_R+v^1\rho\label{sb1'}
\\
S_{b2}^{-1}(\tilde p)&=&i\sum_\mu\gamma_\mu\tilde p_\mu Z_2(\tilde p)P_L.
\label{sb2'}
\end{eqnarray}
However, when $g_1\not=0, \rho\not=0$ eq.(\ref{rho}), the normal mode of the
$\psi_R(x)$ is not guaranteed to completely decouple from that of the
$\psi^i_L(x)$. 

Once we go onto the line A: 
\begin{equation}
g_1=0,\hskip0.5cm g_2^{c,a}<g_2<g_2^{c,\infty},
\label{segment}
\end{equation} 
as indicated
in fig.1, the spectrum (\ref{sn1},\ref{sc1}) is undoubled for $g_2>g_2^{ca}$.
As the results of
the $\psi_R$-shift-symmetry of the action (\ref{action}):
\begin{itemize}
\begin{enumerate}
\item the normal mode
of the $\psi_R$ is a free mode (see eq.(\ref{free}));
\item the NJL mass term 
$\Sigma^1(0)=0$ (see eq.(\ref{ws2}) also eq.(\ref{o}))
for which the $SU_L(2)\otimes U_R(1)$-chiral symmetry is completely restored;
\item the interacting vertex (\ref{4p})
$\Gamma^{(4)}=4g_2w^2(\tilde p)\ll 1$ for the normal modes, which prevent the
normal modes of the $\psi^i_L$ and $\psi_R$ from binding up bound states
$(\bar\psi^i_L\cdot\psi_R)\psi^i_L$, $(\bar\psi_R\cdot\psi^i_L)\psi_R$. 
\end{enumerate}
\end{itemize}
The last point is the most weak point since we base on the discussions of the
wedges opening up due to the momentum-dependent interacting vertex in section
4, rather than calculate the spectrum of normal modes directly. We expect the
last point to be true in a certain segment of the region (\ref{segment}). Thus,
we speculate that there is a possible scaling window for continuum chiral
fermions opening up in this segment. In this possible scaling region, the
spectrum consists of the doublers eq.(\ref{sn1},\ref{sc1}) for $p=\tilde
p+\pi_A$ and the massless normal modes eqs.(\ref{sb1'},\ref{sb2'}) for $g_1=0$, 
\begin{equation}
S^{-1}_L(\tilde p)^{ij}=i\gamma_\mu\tilde p^\mu\tilde Z_2\delta_{ij}P_L;
\hskip1cm S^{-1}_R(\tilde p)=i\gamma_\mu\tilde p^\mu P_R,
\label{sf}
\end{equation}
which are in agreement with the $SU_L(2)\otimes U_R(1)$ symmetry. Namely, this
normal mode of the $\psi_L^i$ is self-scattering via the multifermion coupling
$g_2$ without pairing up with any other modes. The wave function
renormalization $\tilde Z_2$ can be considered as an interpolating constant of
$Z_2(p)$ for $p=\tilde p\simeq 0$ and $g_1=0$. 

If this scenario is truly emerged, in this possible scaling region for the
long distance, we have the massive spectrum that contains fifteen doublers of
the $SU_L(2)$-invariant and $U_R(1)$-covariant neutral Dirac mode $\Psi_n$
eq.(\ref{sn1})($p\not=\tilde p$) and fifteen doublers of the $U_R(1)$-invariant
and $SU_L(2)$-covariant charged Dirac mode $\Psi^i_c$
eq.(\ref{sc1})($p\not=\tilde p$), as well as the $SU_L(2)\otimes U_R(1)$
covariant massive scalar ${\cal A}^i$ eq.(\ref{mas}). Besides, we have massless
spectrum that contains the $U_R(1)$-covariant Weyl mode $\psi_R$ and the
$SU_L(2)$-covariant Weyl mode $\psi^i_L$ eq.(\ref{sf}). In order
to see all possible interactions between these modes in this possible scaling
region, we consider the one-particle irreducible vertex functions of these
modes. In the light of the exact $SU_L(2)\otimes U_R(1)$ chiral symmetry and
$\psi_R$-shift-symmetry, one can straightforwardly obtain non-vanishing vertex
functions ($d$=dimensions) at physical momenta ($p=\tilde p, q=\tilde q$): (i)
${\cal A}^j{\cal A}^{j\dagger}{\cal A}^i{\cal A}^{i\dagger}$ ($d=4$); (ii)
$\bar\psi^i_L\psi_L^i{\cal A}^j{\cal A}^{j\dagger}$, $\bar\Psi^i_c\Psi_c^i{\cal
A}^j{\cal A}^{j\dagger}$ and $\bar\Psi_n\Psi_n{\cal A}^j{\cal A}^{j\dagger}$
($d=5$), as well as $d>5$ vertex functions. The vertex functions with
dimensions $d>4$ vanish in the scaling region as $O(a^{d-4})$ and we are left
with the self-interacting vertex ${\cal A}^j{\cal A}^{j\dagger}{\cal A}^i{\cal
A}^{i\dagger}$. 

In this possible scaling region, the chiral continuum limit is very much like
that of lattice QCD. We need to tune only one coupling $g_1\rightarrow0$ in the
neighborhood of the possible scaling region (\ref{segment}). For $g_1\rightarrow 0$, the
$\psi_R$-shift-symmetry is slightly violated, the normal modes of the
$\psi^i_L$ and $\psi_R$ would couple together to form the chiral symmetry
breaking term $\Sigma^i(0)\bar\psi^i_L\psi_R$, which is a dimension-3
renormalized operator and thus irrelevant at the short distance. We desire this
scaling region to be ultra-violet stable, in which the multifermion coupling
$g_1$ turns out to be an effective renormalized dimension-4 operator\cite{bar}.

\vskip0.7cm
\noindent
{\bf 6.}\hspace*{0.3cm} 
The conclusion of the existence of the possible scaling region (\ref{segment})
for the continuum chiral theory is plausible and hard to be excluded. It is
worthwhile to check and confirm this scenario in different approaches. Even
though, we are still left with several problems. Their possible resolutions are
mentioned and discussed in this section, and deserve to be studied in future
work. 

The question is whether this chiral continuum theory in the scaling region
could be the correct chiral gauge theory, as the $SU(2)$-chiral gauge coupling
$g$ perturbatively is turned on in the theory (\ref{action}). One should expect
a slight change of critical lines (points). We should be able to re-tune the
multifermion couplings ($g_1,g_2$) to compensate these perturbative changes, due
the fact that the gauge interaction does not spoil the $\psi_R$-shift-symmetry
and we have Ward identities
\begin{equation}
{\delta^{(2)} \Gamma\over\delta
A'_\mu\delta\bar\psi'_R}={\delta^{(3)} \Gamma\over\delta
A'_\mu\delta\psi'_R\delta\bar\psi'_R}={\delta^{(3)} \Gamma\over\delta
A'_\mu\delta\Psi'^n_L\delta\bar\psi'_R}=\cdot\cdot\cdot=0,
\nonumber
\end{equation}
where $A'_\mu$ is a ``prime'' gauge field.
In this possible scaling regime, disregarding those uninteresting neutral
modes, we have the charged modes including both the $SU(2)$-chiral-gauged,
massless normal mode (\ref{sf}) of the $\psi^i_L$ and the
$SU(2)$-vectorial-gauged, massive doublers of the Dirac fermion $\Psi^i_c$
(\ref{sc1}), which is made by the 15 doublers of the $\psi^i_L$ and the 15
doublers of the bound Weyl fermion $(\bar\psi_R\cdot\psi^i_L)\psi_R$. The gauge
field should not only chirally couple to the massless normal mode of the
$\psi_L^i$ in the low-energy regime, but also vectorially couple to the massive
doublers of Dirac fermion $\Psi_c^i$ in the high-energy regime. Thus, we expect
the coupling vertex of the $SU_L(2)$-gauge field and the normal mode of the
$\psi^i_L$ to be chiral at the continuum limit. We are supposed to be able to
demonstrate this point on the basis of the Ward identities associating with the
$SU(2)$-chiral gauge symmetry that is respected by the spectrum in the possible
scaling regime. In fact, due to the reinstating of the manifest
$SU_L(2)$-chiral gauge symmetry and corresponding Ward identities of the
undoubled spectrum in this possible scaling regime, we should then apply the
Rome approach\cite{rome} (which is based on the conventional wisdom of quantum
field theory) to perturbation theory in the small gauge coupling. It is
expected that the Rome approach would work in the same way but all
gauge-variant counterterms are prohibited; the gauge boson masses vanish to all
orders of gauge coupling perturbation theory for $g_1=0$. 

Another important question remaining is how chiral gauge anomalies emerge,
although in this short report the chiral gauge anomaly is cancelled by
purposely choosing an appropriate fermion representation of the $SU_L(2)$
chiral gauge group. We know that in the doubled spectrum of naive lattice
chiral gauge theory, the reason for the correct anomaly disappearing in the
continuum limit is that the normal mode and doublers of Weyl fermion produce
the same anomaly these anomalies eliminate themselves\cite{smit}. As a
consequence of decoupled doublers being given chiral-invariant mass $(\sim
O({1\over a}))$, the survival normal mode of the Weyl fermion (chiral-gauged,
e.g., $U_L(1)$) should produce the correct anomaly in the continuum limit. We
also have the question of whether the conservation of fermion number would be
violated by the correct anomaly\cite{ep,bank} structure $\tr F\tilde F$ that is
generated by the $SU(2)$ instanton in the continuum limit. 

I thank G.~Preparata, M.~Creutz, H.B.~Nielsen,
M.~Testa, Y.~Shamir, D.N.~Petcher and M.~Golterman for discussions. 
The auther gratefully acknowledges the support of K.C.~Kong education
foundation, Hong Kong and the national fundation of science, China.

\section*{Figure Captions} 
 
\noindent {\bf Figure 1}: \hspace*{0.2cm} 
The phase diagram for the theory (\ref{action}) in the $g_1-g_2$
plane (at the gauge coupling $g=0$).


\begin{thebibliography}{99}
\bibitem{we}
These calculations are similar to that in ref.\cite{ep} and the details will be
reported elsewhere.
\end{thebibliography}
\end{document}